The Impact of Local Strain Fields in Non-Collinear Antiferromagnetic Films

*Freya Johnson[1,a]\*, Frederic Rendell-Bhatti [2,a], Bryan D. Esser[3], Aisling Hussey[4], David W. McComb[5], Jan Zemen[6], David Boldrin[2], Lesley Cohen[7]*

[1] *Cavendish Laboratory, University of Cambridge, Cambridge, UK*
[2] *SUPA, School of Physics and Astronomy, University of Glasgow, Glasgow G12 8QQ, UK*
[3] *Monash Centre for Electron Microscopy, Monash University, Melbourne, 3800, Australia*
[4] *School of Physics, CRANN and AMBER, Trinity College Dublin, D02PN40, Dublin, Ireland*
[5] *Center for Electron Microscopy and Analysis, The Ohio State University, Columbus, OH, 43212, USA*
[6] *Czech Technical University in Prague, Faculty of Electrical Engineering, Czech Republic*
[7] *Department of Physics, Blackett Laboratory, Imperial College London, London SW7 2AZ, U.K.*
[a]*joint first authors*
*\*Corresponding author*

Abstract

Antiferromagnets hosting structural or magnetic order that breaks time reversal symmetry are of increasing interest for 'beyond von Neumann computing' applications because the topology of their band structure allows for intrinsic physical properties, exploitable in integrated memory and logic function. One such group are the non-collinear antiferromagnets. Essential for domain manipulation is the existence of small net moments found routinely when the material is synthesised in thin film form and attributed to symmetry-breaking caused by spin canting, either from the Dzyaloshinskii–Moriya interaction or from strain. Although the spin arrangement of these materials makes them highly sensitive to strain, there is little understanding about the influence of local strain fields caused by lattice defects on global properties, such as magnetisation and anomalous Hall effect. This premise is investigated by examining non-collinear films that are either highly lattice mismatched or closely matched to their substrate. In either case, edge dislocation networks are generated and for the former case these extend throughout the entire film thickness, creating large local strain fields. These strain fields allow for finite intrinsic magnetisation in seemly structurally relaxed films and influence the antiferromagnetic domain state and the intrinsic anomalous Hall effect.

Introduction

There has been recent great interest in antiferromagnetic materials for their use in energy efficient computing [1-4] and neuromorphic hardware [5,6], with the potential for terahertz operating frequencies, lower thermal losses and enhanced on-chip packing density. [7-10] In particular, non-collinear antiferromagnets (AFM) such as $Mn_3A$ (A = Sn, Ge, Pt) and $Mn_3AN$ (A = Ni, Ga, Sn) have been show to possess intrinsic anomalous Hall effect (AHE) [11-17], magneto-optical Kerr effect (MOKE) [18-20], anomalous Nernst effect (ANE) [21-23] and the tunnelling magnetoresistance effect [24] which allows for easy read-out of the antiferromagnetic state, a hindrance for many collinear antiferromagnetic families [25-27] although encouraging progress with altermagnets may help overcome this bottleneck [28].

Although it is theoretically understood that the AHE may be observed in fully compensated non-collinear AFM with *M* = 0, [29-32] experimentally the observation of anomalous Hall effect in non-collinear AFM systems is accompanied with a small net moment. This moment has been attributed to several sources. For example, in bulk $Mn_3Sn$ single crystals, it is proposed to be a Dzyaloshinskii–Moriya interaction leading to a canting of the spins; [33] in $Mn_3NiN$ thin films grown on closely lattice-matched substrates, the net tetragonal distortion of the lattice is claimed to create a resultant piezomagnetic moment; [34-36] and a recent report on $Mn_3SnN$ thin films grown on mismatched $SrTiO_3$ (STO) substrates show that the global cubic crystal symmetry masks local Mn displacements which creates local symmetry breaking, proposed as the possible source. [37] In all cases, the net moment is coupled to the AFM ordering and is believed to be an essential requirement that allows the AFM domains to be saturated in an applied magnetic field. Were this not the case, all the quantities of interest (AHE, ANE, MOKE) would vanish when averaged out over multiple domains.

Further to this situation, the presence of a weak net moment may confuse the true origin of the AHE, as the association of the AHE in ferromagnets is predicated on a magnetic moment and spin-orbit coupling (SOC). For non-collinear systems moment is not necessarily required, but if a moment is present the origin of the AHE may be obfuscated. Such a question is particularly pertinent in the antiperovskite $Mn_3AN$ and cubic $Mn_3A$ materials, which predominantly adopt one of two non-collinear antiferromagnetic structures - the $\Gamma^{4g}$ structure, which shows anomalous Hall effect, and the $\Gamma^{5g}$ structure, which does not. Presence of the AHE is frequently used to distinguish between these two structures ([11,34,35]), meaning attributing its origin correctly is important.

Within this context, we choose to investigate epitaxial thin films of the non-collinear AFM $Mn_3AN$ family with either large or small lattice-substrate mismatch, and examine how the global properties, AHE and magnetisation, are influenced by atomic scale deformations. Firstly, $Mn_3Cu_{0.5}Sn_{0.5}N$ (MCSN) grown on mismatched MgO substrates, with fully relaxed cubic crystal structure determined globally by X-ray diffraction (XRD) and a Néel temperature of 325K +/- 5K. Secondly, $Mn_3NiN$ (MNN) grown on closely lattice matched STO substrates, with a tetragonal distortion c/a = 1.028 and a Neel temperature of 215 K. [11] All films support an enhancement of AHE at room temperature. In each case, transmission electron microscopy reveals a network of edge dislocations within the films, which produce intense local deformations and strain fields. Edge dislocations are confined to a region 10s of nm from the interface for closely lattice matched films, whilst for the highly mismatched case the dislocation network extends through the entire 70 nm thickness of the film. We demonstrate that although local strain fields are large, their contribution to global strain is vanishingly small when averaged over the entire film, and by considering the interactions of these deformations with the antiferromagnetic structure we show that an enhanced piezomagnetic moment may still be produced. But despite this enhancement, these dislocations are still ultimately detrimental to the intrinsic properties of interest, such as AHE. This work highlights the need for further investigations into the nanoscale properties of these materials, with particular scrutiny of the properties of the interface with disparate materials, where they are being frequently employed such as in tunnel junction applications [38].

<u>Strain field mapping using scanning transmission electron microscopy</u>

For the case of MCSN grown on MgO, it may be anticipated that the large misfit (~6%) will give rise to strain relaxation via plastic deformation and dislocations within a few nm of the interface. [39] In addition, it has been shown that this system demonstrates large magnetovolume effects that manifest as a region of negative thermal expansion on cooling [40] and it can be expected that this will also enhance the internal strain of the sample, due to mismatch between the thermal expansion of the film and the substrate. [39] In Figure 1, we demonstrate that dislocations are indeed present. Figure 1 (a) shows a schematic of an edge dislocation, a real space topological defect where an extra half-plane of atoms is present in the material. Figure 1 (b) shows an image of the interfacial region taken using scanning transmission electron microscopy (STEM). Electron energy loss spectroscopy (EELS) identifies a 3 nm thick mixed-phase interface (1 (c,d)). Although not immediately apparent in the STEM image (1b), the appearance of the dislocations is made clear in Bragg filtered images (1f,g), generated by masking 100 and 001 spots in the FFT (1e)). Using geometric phase analysis (GPA) (see methods), we are able to extract the strain fields from this image in the in-plane direction (1(h)) and in the out-of-plane direction (1(i)) by comparison to the substrate. Within the first 3 nm, the film is closely matched to the substrate in-plane, but sees an expansion out-of-plane, indicating that strain is accommodated elastically within the unit cell. Averaging over rows of atoms (Figure 1 j), this distention is clearly visible, and was present in all areas of the interface examined. Beyond this region, a complex network of edge dislocations develops, producing large strain fields and alternating regions of positive and negative strain, both in- and out-of-plane.

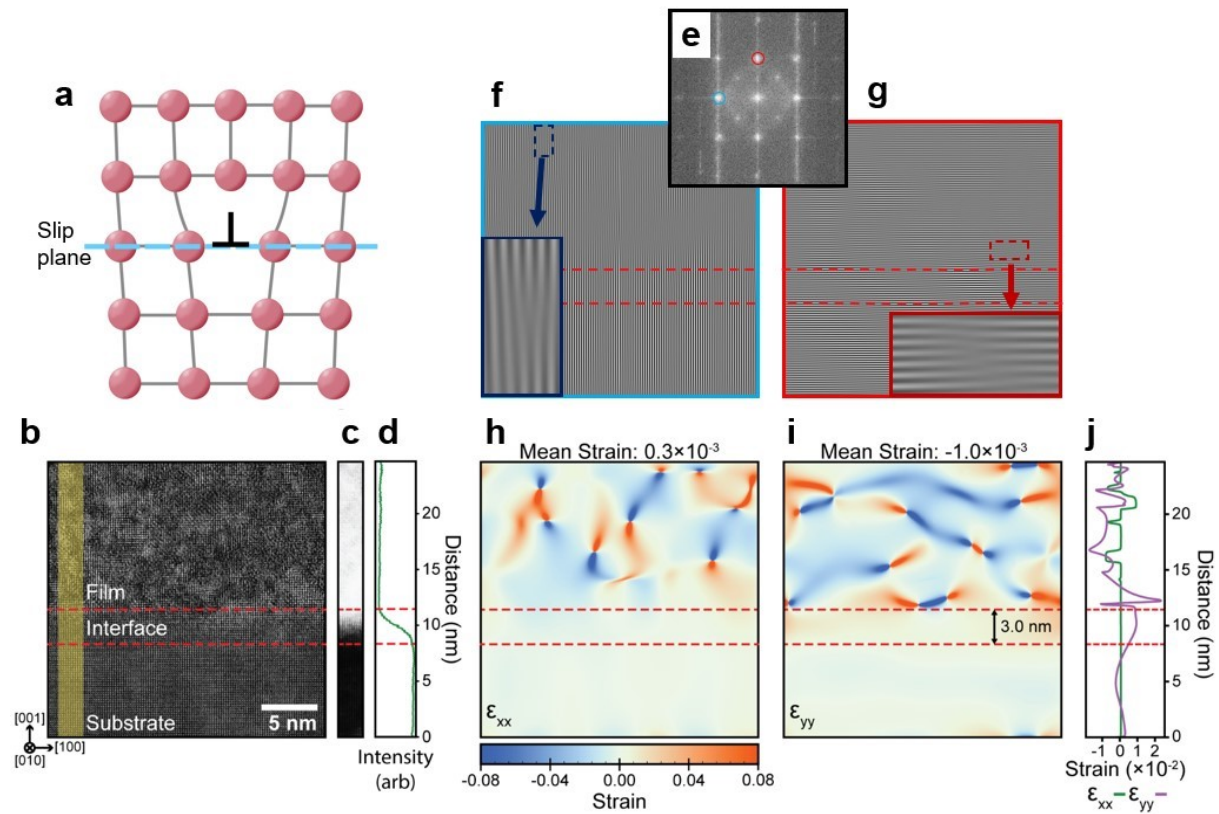

Figure 1. a) Schematic of an edge dislocation, with the slip plane highlighted. The edge dislocation is a defect in the crystal lattice where an extra half plane of atoms is introduced. b) High resolution transmission electron microscopy image of the MCSN film in the interfacial region with the mismatched MgO substrate. c) High-angle annular dark-field image of highlighted region in b). d) Line profile plot of intensity extracted from c), with a 3 nm wide interface clearly defined. e) Power spectrum of image b). The frequencies circled in blue and red are masked and an inverse Fourier transform performed to produce images f) and g) respectively. f,g) 100 and 001 Bragg filtered images. Insets magnify a region featuring an edge dislocation, where the extra half-plane is clearly visible. h,i) Strain maps of the $\varepsilon_{xx}$ and $\varepsilon_{yy}$ components of the strain tensor. The maps reveal a network of dislocations within the film, resulting in large internal strain fields. j) $\varepsilon_{xx}$ and $\varepsilon_{yy}$ averaged over rows of atoms and plotted as a function of distance from the interface.

In order to further understand the strain distribution we sequentially imaged the entire thickness of the film, moving from the interface to the surface, (Figure 2 (a)). From these images, it is clear that the dislocation network extends throughout the film thickness. After averaging the large local strain fields that reach up to 15% at the dislocation core (Figure 2 (c)) the residual behaviour is a small net tetragonality, c/a, that relaxes back to cubic after 35 nm (Figure 2 (b)). Over the entire image series, the mean tetragonality is 1.006. Using X-ray diffraction (XRD), we extract the lattice parameters of the film $a$ = 3.965 +/- 0.005 Å and $c$ = 3.969 +/- 0.002 Å (Figure 2 (d-f)), which shows a close match with the lattice parameters extracted from XRD of MCSN powder $a$ = 3.970 +/- 0.001 Å. This would ordinarily suggest that the MCSN film is relaxed, and the unit cell is cubic within error. Typically a cubic unit cell forbids a piezomagnetic origin of the magnetisation.

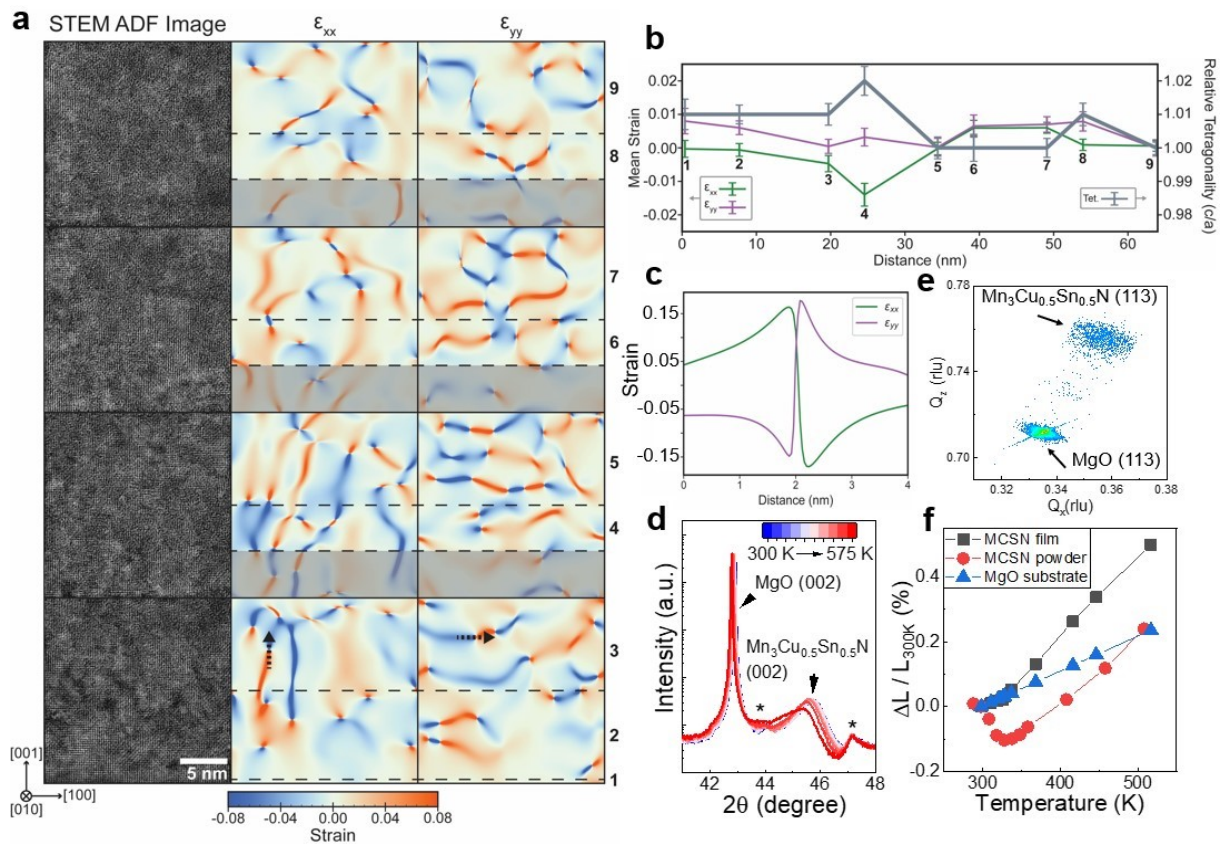

Figure 2. a) Cross-sectional transmission electron microscopy images and corresponding strain fields, moving from the interface towards the surface of the film. Edge dislocations are present throughout all regions of the film. b) Average strain and tetragonality as a function of distance from the interface. Numbers 1 – 9 correspond to the regions highlighted in a). Within 35 nm of the interface, the film possesses a small average tetragonality, c/a > 1. c) Line scan over the region in a) marked by arrows. Cloe to the dislocation, the strain reaches a maximum value of 15%. d) X-ray diffraction of the 002 reflection as a function of temperature, demonstrating epitaxy between the film and the substrate in agreement with TEM. At 575 K, the highest temperature probed, the film shows signs of oxidation. Small peaks marked with an asterisk are due to the heater stage. e) Reciprocal space map around the 113 and 002 reflection respectively, from which we extract a = 3.965 +/- 0.005 Å and c = 3.969 +/- 0.002 Å. As X-ray diffraction samples properties over the entire film, the film appears cubic within error. (f) Relative change in c lattice parameter as a function of temperature. In the film, a change in the coefficient of thermal expansion is observed at $T_N$ = 325 +/- 5 K, in agreement with powder diffraction data, although no region of negative

thermal expansion is observed. The presence of the magnetovolume effect in this composition is a contributing factor to the high density of dislocations.

We now turn to the closely matched sample, MNN on STO, with misfit of 0.6% (Figure 3). Similar to the case of MCSN, immediately beyond the interface region a network of dislocations develops, but unlike in MCSN the dislocation density is greatly reduced after a critical thickness of ~20 nm. Images taken further away from the substrate (see supplementary) suggest the film grows relatively defect-free, but with a constant tetragonality c/a = 1.028. This allows for a piezomagnetic moment and therefore control of the antiferromagnetic domain structure with applied field, as we have previously reported. [11,13]

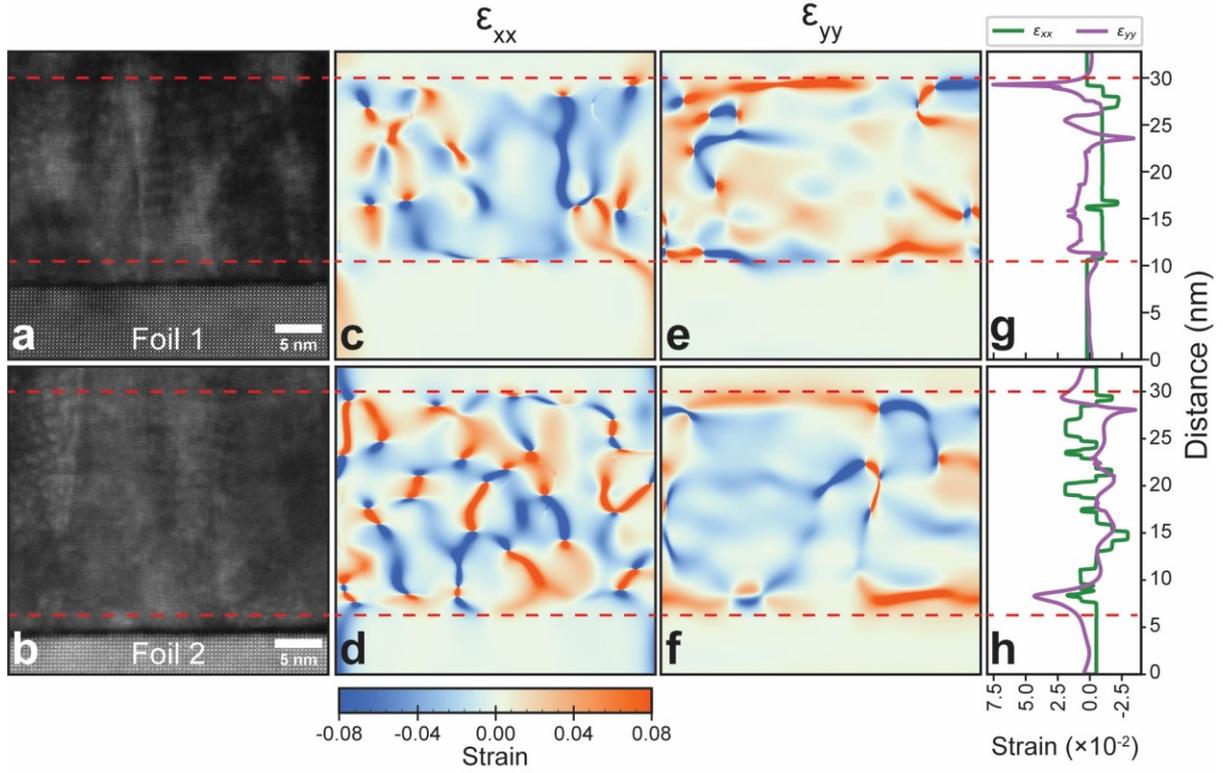

Figure 3. a,b) High resolution transmission electron microscopy images from two foils from a MNN film grown on STO substrate. c-f) Strain maps of the $\varepsilon_{xx}$ and $\varepsilon_{yy}$ components of the strain tensor. As in Figure 1, beyond the interface a network of dislocations develops, but above 20 nm the dislocation density is reduced. g,h) $\varepsilon_{xx}$ and $\varepsilon_{yy}$ averaged over rows of atoms and plotted as a function of distance from the interface.

Magnetisation and Magnetotransport measurements

To understand the influence of the internal strain on the magnetic properties, we focus on the magnetisation and magnetotransport of the highly mismatched MCSN film. In ferromagnetic materials, the Hall resistivity, $\rho_H$, may be written as the sum of the ordinary Hall resistivity $\rho_{OH}$ (linear in magnetic flux density $B$) and the anomalous Hall resistivity $\rho_{AH}$, proportional to magnetisation $M$:

$$\rho_H = \rho_{OH} + \rho_{AH} = R_0 B + R_s \mu_0 M \qquad (1)$$

Where $R_0$ is the ordinary Hall coefficient and $R_s$ is the extraordinary Hall coefficient. $\rho_{AH}$ may further be divided into terms proportional to longitudinal resistivity $\rho_{xx}$ and $\rho_{xx}^2$:

$$\rho_{AH} = a(M)\rho_{xx} + (b(M) + c)\rho_{xx}^2 \qquad (2)$$

Or, in terms of the conductivity:

$$\sigma_{xy} = a(M)\sigma_{xx} + b(M) + c \quad (3)$$

Where we have explicitly introduced the constant $c$ to represent an intrinsic contribution that is independent of the magnetisation. The term proportional to $\sigma_{xx}$ represents the skew-scattering contribution which dominates in the "clean limit", while the other terms represent the intrinsic contributions due to the Berry curvature of the band structure, and dominate within the intrinsic regime. [41,42]

In a multidomain sample, the presence of an net moment intrinsic to the film (but independent of its origin) is essential to control the domain orientation, meaning that the coercive field (the field at which all domains are aligned to the applied field) is common to both the magnetization and the intrinsic AHE. Indeed in Figure 4 (a,b) we see good agreement between the $M(H)$ and AHC curves, once a soft temperature independent component of $M(H)$ is removed (Figure 4 (c)) (see supplementary). Because of this shared origin, plotting AHE versus M at 300 K yields the expected linear relationship (Figure 4 (d)), and the gradient may then be used to extract $R_s$, according to equation 1. From this, we find $R_s$ = 1.01E-7 m$^3$C$^{-1}$ at 300 K, two orders of magnitude larger than in ferromagnets such as Fe, Co and Ni. [43]. This suggests that the antiferromagnetic structure is the dominant contribution to the AHE, beyond the contributions of the weak net moment, and is indicative of $\Gamma^{4g}$. In addition, Figure 4 (e) demonstrates that $\sigma_{xy}$ is independent of $\sigma_{xx}$ and from equation (3) the sample lies within the intrinsic AHE regime.

Additional evidence that the AFM structure is $\Gamma^{4g}$ comes from the temperature dependence of the ratio of the saturated magnetisation in-plane to out-of-plane, which approaches $1/\sqrt{2}$, the anticipated ratio for a piezomagnetic moment in the [112] direction produced from tetragonally distorted $\Gamma^{4g}$ (Figure 4 (f)). [34] The presence of moment out-of-plane excludes tetragonally distorted $\Gamma^{5g}$, which only presents a magnetisation in-plane. [44] Finally, further indications that the AHE origin is not directly related to the magnetisation can be seen by examining the temperature dependence of both. In Figure 4 (g), $M_{sat}$ steadily increases from 340 K to 300 K, but does not increase any more upon further cooling. In contrast, the AHE demonstrates only a small increase until 310 K, whereupon it rapidly increases.

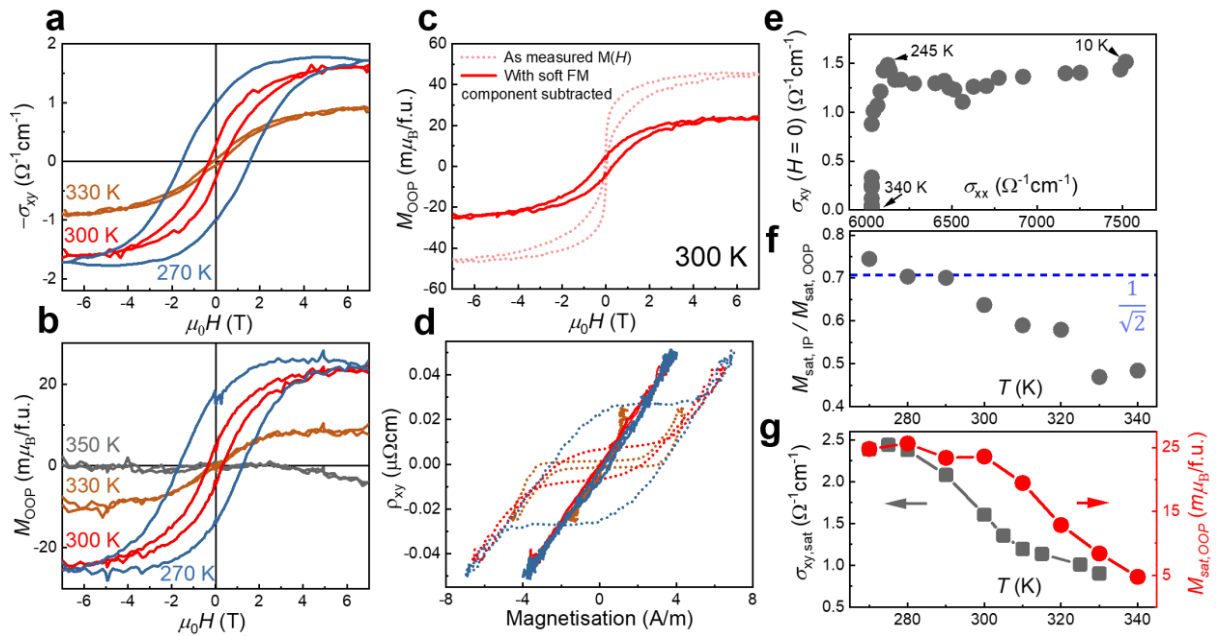

Figure 4. a) The anomalous Hall conductivity at selected temperatures, indicative of the $\Gamma^{4g}$ magnetic phase. b) Out-of-plane M(H) at selected temperatures, with soft ferromagnetic component subtracted. c) Out-of-plane M(H) at 300 K. The dashed curve indicates the signal measured after subtraction of the diamagnetic background. The solid curve is the resultant signal after further subtraction of a soft ferromagnetic component (see supplementary material). d) Anomalous Hall resistivity vs magnetisation. The dashed curves are produced using M(H) as measured, while the solid curve shows the linear relationship once the soft ferromagnetic component is subtracted. e) Remnant $\sigma_{xy}$ vs $\sigma_{xx}$ demonstrating the sample lies in the intrinsic regime. f) Temperature dependence of the ratio of in-plane to out-of-plane saturated magnetisation. On cooling, the ratio approaches $1/\sqrt{2}$, the anticipated ratio for the piezomagnetic moment produced from tetragonally distorted $\Gamma^{4g}$. g) Temperature dependence of saturated $\sigma_{xy}$ and out-of-plane magnetization. The observed temperature dependencies demonstrate that $\sigma_{xy}$ cannot be said to be caused by the magnetisation.

Discussion

The large strain fields from the dislocations will produce piezomagnetic moments. [34-36,45] The global properties we measure, AHE and magnetisation, will be a result of the interactions between the antiferromagnetic domain structure, the localised strain fields and the net moments. As we will now explain, the impact of dislocations, while producing a magnetisation, could be deleterious to the magnitude of the measured anomalous Hall effect.

In various studies of antiperovskite thin films [11,34,37,46] it has been consistently observed that the measured magnetisation is larger than that anticipated from the net tetragonality (as observed by XRD) and the theoretically predicted magnitude of piezomagnetism. These observations are illustrated in Figure 5 (a). But as XRD is a global measurement, it is not sensitive to the local tetragonal distortions produced by edge dislocations, which present in the form of a "butterfly" shaped region surrounding the defect where c/a > 1 above the slip plane, and c/a < 1 below it (Figure 5 (b-e)). It has also commonly been observed that the M(H) loops possess a soft component that is not present in the AHE(H) loops, as is also the case here. Now considering the AFM structure, the $\Gamma^{4g}$ structure has eight possible variants, corresponding to the 8 (111) planes. [21] Each variant may, under a tetragonal distortion, produce a piezomagnetic moment oriented in one of the eight <112> directions, with the direction of the moment determined by the local $\Gamma^{4g}$ variant (Figure 5 (f)). However, for a single $\Gamma^{4g}$ variant, moving from *c* > a to *c* < a reverses the direction of the net moment (Figure 5 (g,h)).

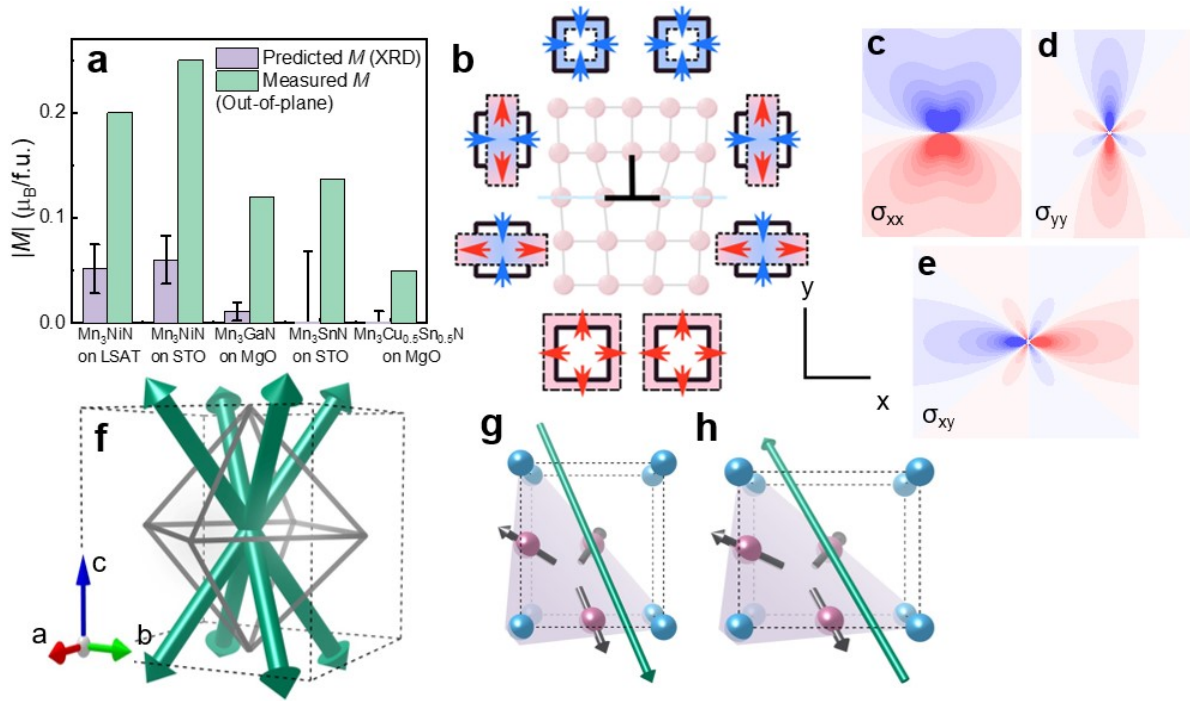

Figure 5. a) Comparison of theoretically predicted piezomagnetic magnetisation, using tetragonality measured via X-ray diffraction, to experimentally measured values out-of-plane. Data are from ref. [11, 37, 46] and this work. b) Schematic of tetragonal distortions created from the stress field of an edge dislocation. c,d,e) Components of the stress tensor surrounding the dislocation. f) Illustration of the possible piezomagnetic moments produced from tetragonal strain. The non-collinear AFM structure may be arranged in any of the eight {111}-type planes (highlighted in grey) and each arrangement produces a unique moment in one of the eight equivalent <112> directions. This feature allows saturation of the domain state with a magnetic field applied in an appropriate direction. g,h) The piezomagnetic moment for a fixed antiferromagnetic domain state may be reversed by reversing the tetragonality i.e. going from c/a > 1 to c/a <1. Regions of mixed tetragonality produced by edge dislocations therefore may impact the magnetic properties of the film.

To understand the implications of this we first consider the case of a single dislocation (Figure 6). Figure 6 (a) shows the simplest case, where the antiferromagnetic ordering surrounding the defect is constant, which we label Type 1. In this instance, the tetragonal regions above the slip plane produce a net moment in the opposite direction to the regions below the slip plane, similar to a ferromagnetic domain wall, but the sign of the intrinsic AHE is the same in all regions. These counteracting net moments prevent control of the antiferromagnetic order by magnetic field, creating frozen regions within the film. A heavily defected sample will therefore show reduced AHE and M, as the lightly strained regions between defects are the only part that is sampled when sweeping magnetic field, greatly reducing the active volume of the film. Experimentally, the presence of exchange bias in films of $Mn_3Sn$ has been explained using frozen antiferromagnetic domains, which also demonstrates piezomagnetism, although the crystal structure and magnetic symmetries differ in this case.[43]

In the second scenario (Figure 6 b) we consider that regions above and below the slip plane may have opposite antiferromagnetic order, with an antiferromagnetic domain wall between them (Type 2). Now the moments add, and the order may be controlled by applied field, but the domains will always be opposite above and below the slip plane. This leads to cancellations in AHE in the areas surrounding defects, as they are locked into a multidomain state that cannot be changed with applied magnetic field. The step feature present in M, but not in AHE, can now be understood as due to the presence of Type 2 structures. The large distortions in the vicinity of the dislocation produce large piezomagnetic moments, which strongly couple to the applied field, leading to a narrow coercive field in comparison to the rest of the sample.

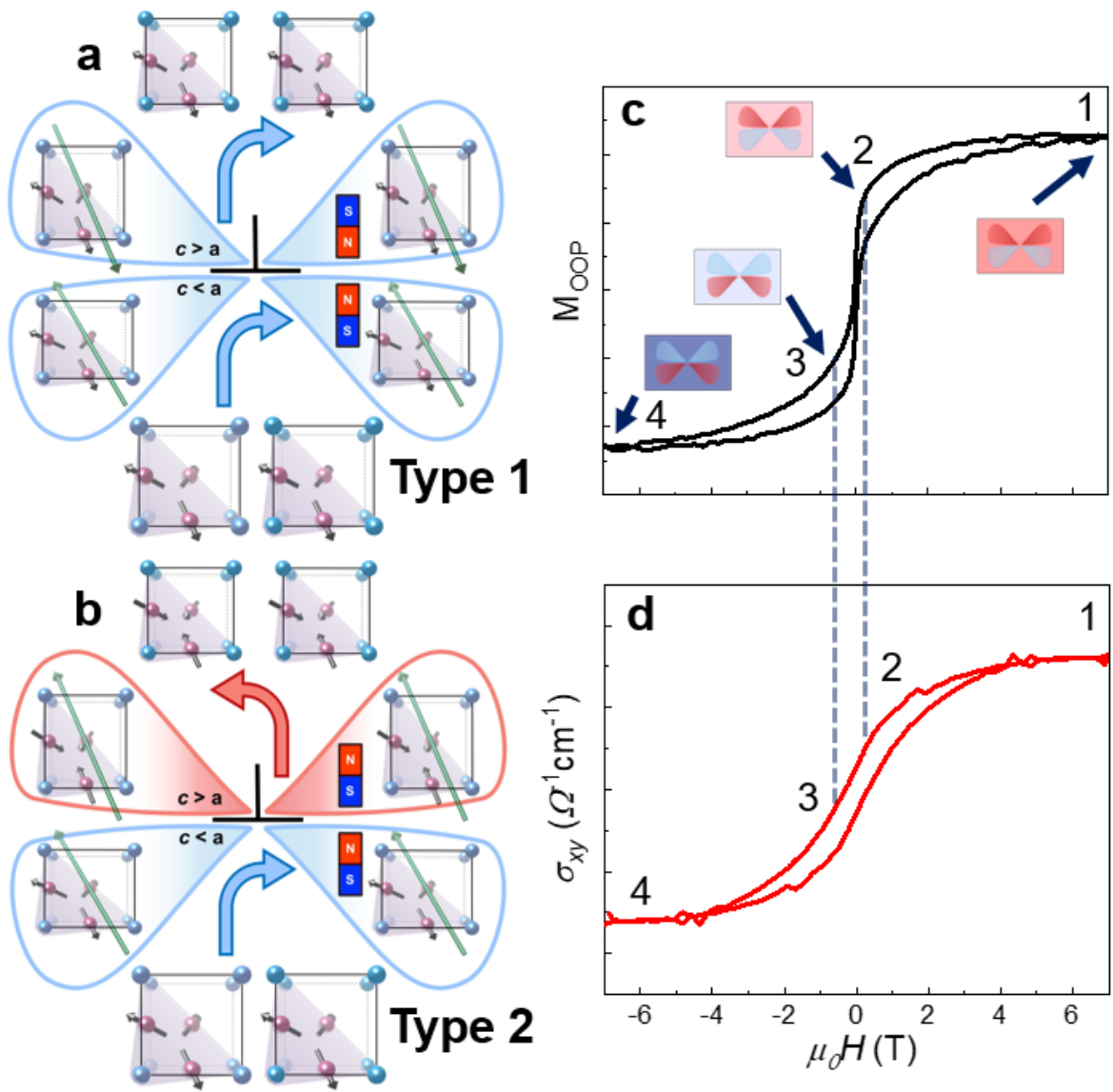

Figure 6. a,b) Schematics demonstrating how the strain surrounding a dislocation, as shown in Figure 4, may influence the magnetic behaviour of the sample. We identify two types: a) Type 1, where the antiferromagnetic domain structure is unaffected by the presence of the edge dislocation. The reversal of tetragonality above and below the slip plane creates a region of opposed magnetic moments that will not respond to an applied field. The structure produces vanishing net moment and cannot be measured in a typical AHE measurement, as the analysis typically requires antisymmetrization of the signal in applied field. This scenario leads to a "frozen region" of the film in the vicinity of the dislocation. b) Type 2, where there is an antiferromagnetic domain wall located in the slip plane of the dislocation. Now, the net

magnetic moments are aligned, and the feature may be switched in magnetic field. Due to the mixed domain state, this structure will not produce a net AHE, but will show a net magnetisation, in contrast to Scenario 1. c,d) M(H) and AHE at 300 K. The step feature present in M, but not in AHE, can now be understood as due to the presence of Type 2 structures. The large distortions in the vicinity of the dislocation produce large piezomagnetic moments, which strongly couple to the applied field, leading to a narrow coercive field in comparison to the rest of the sample. The anomalous Hall conductivity measured is an average behaviour and is non-zero due to the small deviation from cubic tetragonality as determined by the HRTEM strain field analysis.

Both scenarios cause a reduction in the measured intrinsic properties of the film. One can imagine further scenarios where the antiferromagnetic domain wall is present at an arbitrary angle to the slip plane. But the crux of the issue is that it is desirable to have continuous magnetic order across the entire sample, to maximise intrinsic properties, but it is also desirable to have the same piezomagnetically induced local moment across the sample, to facilitate control of the antiferromagnetic order in applied field. Achieving both simultaneously is impossible in a region of mixed tetragonality. For this reason, further understanding of the internal strain, to ensure there is uniform tetragonality throughout the entire sample, is essential. We speculate that the confinement of the dislocations to the interface region in the MNN film is primarily due to the 10x smaller misfit between MNN and $SrTiO_3$ compared to MCSN and MgO, but another contributing factor may be the much smaller magnetovolume effect in MNN compared to MCSN. [35,40] In a recent study on $Mn_3SnN$ on $SrTiO_3$ with intermediate misfit of 3.9% it was observed that Mn atomic displacements observed using highly brilliant grazing incidence XRD reached a maximum for films 50 nm thick, and this was attributed to internal strain. [37] Although the specific details of the internal strain distribution in the film will depend on multiple factors, this result is consistent with our own observations. Taken together, this suggests that by reducing lattice/substrate mismatch, or by introducing buffer layers, or by tuning the $Mn_3AN$ lattice parameter through doping, the dislocation density of these fascinating nitride films can be controlled. In MNN thin films, our measurements of AHE [11,13,16], MOKE [20] and ANE [21,23], while still large in comparison to the weak magnetic moment, are greatly reduced in comparison to theoretical predictions, which may partially result from mixed tetragonality regions within the films, particularly close to the interface as we have shown here.

Conclusions

The properties of non-collinear antiferromagnets are highly sensitive to strain, and this includes large local strain fields generated by dislocations. Although thin films may appear to be relaxed under global XRD techniques, high-resolution TEM and corresponding GPA analysis reveals a network of dislocations originating from misfit strain. In the case of MCSN on MgO, the network extends throughout the entire film thickness. MNN films grown on closely lattice matched STO substrates show a similar region of high dislocation density, but confined to a critical thickness, and beyond this region the film grows with a net tetragonality and few defects. Regions of mixed tetragonality, generated by dislocations, may enhance the magnetisation but will reduce intrinsic quantities of interest due to the interactions between the strain fields and the AFM structure. A future challenge is to investigate techniques to minimize the dislocation density when growing on highly mismatched substrates, while maintaining a constant net tetragonality of the unit cell by identifying appropriate compliant or buffer layers. This analysis will be increasingly important as efforts are made to integrate these materials into multilayer devices. It is clear from this study that these defects, whilst enhancing the local moment associated with piezomagnetism, are in fact deleterious to the physical properties important for application.

Experimental section

Samples were prepared using standard solid state synthesis techniques. First, $Mn_2N$ was formed by reacting elemental Mn powder (Alfa Aesar, 99.95% purity, 325 mesh) under dry, flowing nitrogen gas at 973 K for 48 h. This $Mn_2N$ precursor was first ball-milled using 3 mm ceramic balls at 300 RPM for 1 hour with IPA as the solvent and then thoroughly mixed with elemental Cu (Sigma-Aldrich, 99.7% purity, 3 μm). The resulting mixture was pressed into a ~2 g pellet, wrapped in Ta foil and transferred to a crucible. The crucible was heated at 1053 K for 2 days and allowed to cool to ambient temperatures. This sample, with composition $Mn_3CuN$ as determined by x-ray diffraction, was turned into a powder via ball-milling. The previous technique was repeated with elemental Sn (Alfa Aesar, 99.85% purity, 100 mesh) to form Mn3SnN powder. The two powders, $Mn_3CuN$ and $Mn_3SnN$, were thoroughly combined according to the ratio $Mn_3Cu_{0.5}Sn_{0.5}N$. This mixture was subsequently pressed into a pellet with a 2 cm diameter and heated under dry, flowing nitrogen at 1053 K for 2 days, again allowing the sample to cool to ambient temperatures.

Thin films of $Mn_3Cu_{0.5}Sn_{0.5}N$ were grown by pulsed laser deposition at 500 °C using a KrF excimer laser (λ = 248 nm) with a laser fluency of 0.8 J/cm$^2$, under 5 mTorr nitrogen partial pressure. The bulk lattice parameter of the $Mn_3Cu_{0.5}Sn_{0.5}N$ target is $a$ = 3.970. Four terminal magnetotransport data were collected in a square geometry using the van der Pauw method. Magnetization measurements were performed using a Quantum Design Magnetic Property Measurement System (MPMS 3).

Transmission electron microscopy data were taken using an JEOL ARM 200cF and FEI Titan in STEM mode. Images were obtained using an annular dark-field (ADF) detector (Figs. 1 & 2) and a high-angle annual dark-field (HAADF) detector (Fig. 3).

The geometric phase analysis was performed using the Strain++ software, which contains algorithms for determining strain fields in high-resolution TEM images as detailed in [47,48]. Raw STEM ADF and HAADF images were used for analysis, by selecting two orthogonal spots in the FFT (100 and 001). For images that contained the MgO substrate, refinement of the GPA analysis was performed using a small region of the substrate. For images that did not contain the MgO substrate, a calibration image was taken of the substrate using identical imaging conditions to the subsequent images of the film. In this case, refinement was done using the substrate calibration image.

Acknowledgements

F.J is grateful for support from a Royal Commission of 1851 Research Fellowship, and from the Henry Royce Institute for Advanced Materials.

A.H was supported by Science Foundation Ireland 18/EPSRC-CDT-3581 and the Engineering and Physical Sciences Research Council EP/S023259/1

D B is grateful for support from a Leverhulme Trust Early Career Fellowship (No. ECF-2019-351) and a University of Glasgow Lord Kelvin Adam Smith Fellowship.